\documentclass[twocolumn,showpacs,preprintnumbers,amsmath,amssymb]{revtex4}
\usepackage{graphicx}
\usepackage{dcolumn}
\usepackage{bm}
\usepackage{amsmath}
\usepackage{amsfonts}
\usepackage{amssymb}
\usepackage{setspace}

\def\ket#1{\lvert#1\rangle}
\def\bra#1{\langle #1\lvert}

\def\ketbra#1#2{\lvert#1\rangle\langle#2\lvert}

\def\1{\mathchoice{\rm 1\mskip-4.2mu l}{\rm 1\mskip-4.2mu l}{\rm
        1\mskip-4.6mu l}{\rm 1\mskip-5.2mu l}}

\begin{document}

\title{Geometric Phase in Entangled Systems: A Single-Neutron Interferometer Experiment }
\author{S. Sponar$^1$}
\email {sponar@ati.ac.at}
\author{J. Klepp$^{1}$}
\author{R. Loidl$^1$}
\author{S. Filipp$^2$}
\author{K. Durstberger-Rennhofer$^1$}
\author{R. A. Bertlmann$^3$}
\author{G. Badurek$^1$}
\author{Y. Hasegawa$^{1}$}
\author{H. Rauch$^{1,4}$}
\affiliation{%
$^1$Atominstitut der \"{O}sterreichischen Universit\"{a}ten, 1020 Vienna, Austria\\$^2$ Department of Physics, ETH Zurich, Schafmattstr. 16,
8093 Zurich, Switzerland\\ $^3$ Faculty of Physics, University of Vienna, Boltzmanngasse 5, A-1090 Vienna, Austria\\ $^4$Institut Laue-Langevin,
B.P. 156, F-38042 Grenoble Cedex 9, France}

\date{\today}
\begin{abstract}

The influence of the geometric phase on a Bell measurement, as proposed by Bertlmann $et$ $al.$ in [Phys. Rev. A 69, 032112 (2004)], and
expressed by the Clauser-Horne-Shimony-Holt (CHSH) inequality, has been observed for a spin-path entangled neutron state in an interferometric
setup. It is experimentally demonstrated that the effect of geometric phase can be balanced by a change in Bell angles. The geometric phase is
acquired during a time dependent interaction with two radio-frequency (rf) fields. Two schemes, polar and azimuthal adjustment of the Bell
angles, are realized and analyzed in detail. The former scheme, yields a sinusoidal oscillation of the correlation function $S$, dependent on
the geometric phase, such that it varies in the range between 2 and $2\sqrt{2}$ and, therefore, always exceeds the boundary value 2 between
quantum mechanic and noncontextual theories. The latter scheme results in a constant, maximal violation of the Bell-like-CHSH inequality, where
$S$ remains $2\sqrt2$ for all settings of the geometric phase.

\end{abstract}


\maketitle

\section{\label{sec:level1}Introduction}

Since the famous 1935 Einstein-Podolsky-Rosen (EPR) gedanken experiment \cite{EPR35} much attention has been devoted to quantum entanglement
\cite{schroedinger35}, which is among the most striking peculiarities in quantum mechanics. In 1964, J. S. Bell \cite{Bell64} introduced
inequalities for certain correlations, which hold for the predictions of any hidden-variable theory applied \cite{BookBell}. However, a
dedicated experiment was not feasible at the time. Five years later Clauser, Horne, Shimony and Holt (CHSH) reformulated Bell\char39{}s
inequality (BI) pertinent for the first practical test of the EPR claim \cite{Clauser69}. Thereafter polarization measurements with correlated
photon pairs \cite{BookBertlmannZeilinger}, produced by atomic cascade \cite{Freedman72,Aspect81} and parametric down-conversion of lasers
\cite{Kwiat95,Weihs98,Gisin98}, demonstrated a violation of the CHSH inequality. Up to date several systems
\cite{Rowe01,moehring04,sakai06,Matsukevich2008} have been examined, including neutrons \cite{Hasegawa03Bell}.

EPR experiments are designed in order to  test local hidden variable theories (LHVTs) thereby exploiting the concept of nonlocality. LHVTs are a
subset of a larger class of hidden-variable theories known as the noncontextual hidden-variable theories (NCHVTs). Noncontextuality implies that
the value of a measurement is independent of the experimental context, i.e. of previous or simultaneous measurements \cite{Bell66,Mermin93}.
Noncontextuality is a more stringent demand than locality because it requires mutual independence of the results for commuting observables even
if there is no spacelike separation \cite{Simon00}. First tests of quantum contextuality, based on the  Kochen$-$Specker theorem
\cite{Kochen67}, have been recently proposed \cite{Cabello08,Cabello08n} and performed successfully using trapped ions \cite{Kirchmair09} and
neutrons \cite{Hasegawa2006contextual,Hasegawa2009}.

In the case of neutrons, entanglement is not achieved between different particles, but between different degrees of freedom. Since the
observables of one Hilbert spaces (HS) commute with observables of a different HS, the single neutron system is suitable for studying NCHVTs.
Using neutron interferometry \cite{Rauch74,Rauch00Book} single-particle entanglement, between the spinor and the spatial part of the neutron
wave function \cite{Hasegawa03Bell}, as well as full tomographic state analysis \cite{hasegawa2007tomography}, have already been accomplished.
In addition creation of a triply entangled single neutron state \cite{Sponar07}, by applying a coherent manipulation method of a neutron's
energy degree of freedom, has been demonstrated.

The total phase acquired during an evolution of a quantal system generally consists of two components: the usual dynamical phase $-1/\hbar\int
H(t) dt $, which depends on the dynamical properties, like energy or time, and a geometric phase $\gamma$, which is, considering a spin
$\frac{1}{2}$ system, minus half the solid angle ($\Omega/2$) of the curve traced out in ray space. The peculiarity of this phase, first
discovered by M.\,V.\,Berry in 1984 \cite{Berry84}, lies in the fact that it does not depend on the dynamics of the system, but purely on the
evolution path of the state in parameter space. From its first verification, for photons in 1986 \cite{Tomita86} and later for neutrons
\cite{Bitter87}, generalizations  such as non adiabatic \cite{Aharanov87}, noncyclic \cite{Samuel88}, including the Pancharatnam relative phase
\cite{Pancharatnam56}, off-diagonal evolutions \cite{Manini00,Hasegawa01,Hasegawa02}, as well as the mixed state case
\cite{Sjoeqvist00prl,Filipp03,FilippSjoequist03,Klepp05MixedPancharatnamPhase,Klepp08}, have been established.

The geometric phase in a single-particle system has been studied widely over the past two and a half decades. Nevertheless its effect on
entangled quantum systems is less investigated. The geometric phase is an excellent candidate to be utilized for logic gate operations in
quantum communication \cite{QI1}, due to its robustness against noise, which has been tested recently using superconducting qubits
\cite{Leek07}, and trapped polarized ultracold neutrons \cite{Filipp09}. Entanglement is the basis for quantum communication and quantum
information processing, therefore studies on systems combing both quantum phenomena, the geometric phase and quantum entanglement, are of great
importance \cite{Sjoeqvist00pra,Durstberger04,Tong03}.

This article reports on an experimental confirmation for the violation of Bell's inequality, relying on correlations between the spin and path
degrees of freedom of a single neutron system, under the influence of the geometric phase. The geometric phase is generated in one of the
complementary Hilbert spaces, in our case the spin subspace. We demonstrate in detail how the geometric phase affects the Bell angle settings,
required for a violation of a Bell inequality in the CHSH formalism, in a polarized neutron interferometric experiment. In section
\ref{sec:theory} the theoretical framework, as developed in \cite{Durstberger04}, is briefly described for a spin-path entangled neutron state.
Expectation values and Bell-like inequalities are defined and the concept of \emph{polar} and \emph{azimuthal angle adjustment} is introduced.
Section \ref{sec:experiment} explains the actual measurement process. It focuses on experimental issues such as state preparation, manipulation
of geometric phase, joint measurements, as well as the experimental strategy. In the principal part data analysis and experimental results are
presented. This is followed by sections \ref{sec:discussion} and \ref{sec:conclusion} consisting of discussion, conclusion and acknowledgments.

\section{\label{sec:theory}Theory}

\subsection{Expectation Values}

First we want to clarify the notation, since numerous angles are due to appear in this article. Angles denoted as $\boldsymbol\alpha$ are
associated with path, and angles denoted as $\boldsymbol\beta$ with spin subspace. The $^\prime$ symbol is used to distinguish different
measurement directions of one subspace, required for a CHSH-Bell measurement \cite{Clauser69} (e.g. $\boldsymbol\alpha$ and
$\boldsymbol\alpha^\prime$ represent the measurement directions for the path subspace). Index 1 denotes polar angles, whereas index 2 is
identified with azimuthal angles (e.g. $\beta_1$ and $\beta'_1$ are polar angles of the spin subspace). Finally, the $^\bot$ symbol is used for
adding $\pi$ to an angle (e.g. $\alpha_1^\bot=\alpha_1+\pi$).

Following the notation given in \cite{Durstberger04}, in our experiment the neutron\char39{}s wavefunction is defined in a tensor product of two
Hilbert spaces: One Hilbert space is spanned by two possible paths in the interferometer given by $\ket{\textrm{I}},\ket{\textrm{II}}$, and the
other by spin-up and spin-down eigenstates, denoted as $\ket{\Uparrow}$ and $\ket{\Downarrow}$, referred to a quantization axis induced by a
static magnetic field. Interacting with a time dependent magnetic field, the entangled Bell state acquires a geometric phase $\gamma$ tied to
the evolution within the spin subspace \cite{Durstberger04}
\begin{equation}
\ket{\Psi_{\textrm{Bell}}(\gamma)}=\frac{1}{\sqrt{2}}\Big(\ket{\textrm{I}}\otimes\ket{\Uparrow}+ \ket{\textrm{II}}\otimes
e^{i\gamma}\ket{\Downarrow}\Big).
\end{equation}
As in common Bell experiments a joint measurement for spin and path is performed, thereby applying the projection operators for the path
\begin{equation}
\hat P^{\textrm{p}}_\pm(\boldsymbol{\alpha})=\ketbra{\pm\boldsymbol{\alpha}}{\pm\boldsymbol{\alpha}},
\end{equation}
with
\begin{eqnarray}
\ket{+\boldsymbol\alpha}&=&\cos\frac{\alpha_1}{2}\ket{\textrm{I}}+e^{i\alpha_2}\sin\frac{\alpha_1}{2}\ket{\textrm{II}}\nonumber\\
\ket{-\boldsymbol\alpha}&=&-\sin\frac{\alpha_1}{2}\ket{\textrm{I}}+e^{i\alpha_2}\cos\frac{\alpha_1}{2}\ket{\textrm{II}},
\end{eqnarray}
where $\alpha_1$ denotes the polar angle and $\alpha_2$ the azimuthal angle, and, for the spin subspace,
\begin{equation}
\hat P^s_\pm(\boldsymbol{\beta})=\ketbra{\pm\boldsymbol{\beta}}{\pm\boldsymbol{\beta}},
\end{equation}
with
\begin{eqnarray}
\ket{+\boldsymbol\beta}&=&\cos\frac{\beta_1}{2}\ket{\Uparrow}+e^{i\beta_2}\sin\frac{\beta_1}{2}\ket{\Downarrow}
\nonumber\\
\ket{-\boldsymbol\beta}&=&-\sin\frac{\beta_1}{2}\ket{\Uparrow}+e^{i\beta_2}\cos\frac{\beta_1}{2}\ket{\Downarrow}.
\end{eqnarray}
Introducing the observables
\begin{eqnarray}
    \hat A^{\textrm{p}}(\boldsymbol{\alpha})&=&\hat P_{+}^{\textrm{p}}(\boldsymbol{\alpha})-\hat P_{-}^{\textrm{p}}(\boldsymbol{\alpha})\nonumber\\
    \hat B^{\textrm{s}}(\boldsymbol{\beta})&=&\hat P_{+}^{\textrm{s}}(\boldsymbol{\beta})-\hat P_{-}^{\textrm{s}}(\boldsymbol{\beta})\
\end{eqnarray}
one can define an expectation value for a joint measurement of spin and path along the directions $\boldsymbol\alpha$ and $\boldsymbol\beta$
\begin{eqnarray}\label{eq:jointMeasurement}
\begin{split}
&E(\boldsymbol\alpha,\boldsymbol\beta)=\bra{\Psi} \hat A^{\textrm{p}}(\boldsymbol\alpha)\otimes \hat B^{\textrm{s}}(\boldsymbol\beta)\lvert\Psi\rangle\\
&=-\cos\alpha_1\cos\beta_1-\cos(\alpha_2-\beta_2+\gamma)\sin\alpha_1\sin\beta_1 \\
  &=-\cos(\alpha_1-\beta_1)\textrm{ for } (\alpha_2-\beta_2)= -\gamma.
\end{split}
\end{eqnarray}

%
%
\begin{figure*}
{\includegraphics [width=180mm] {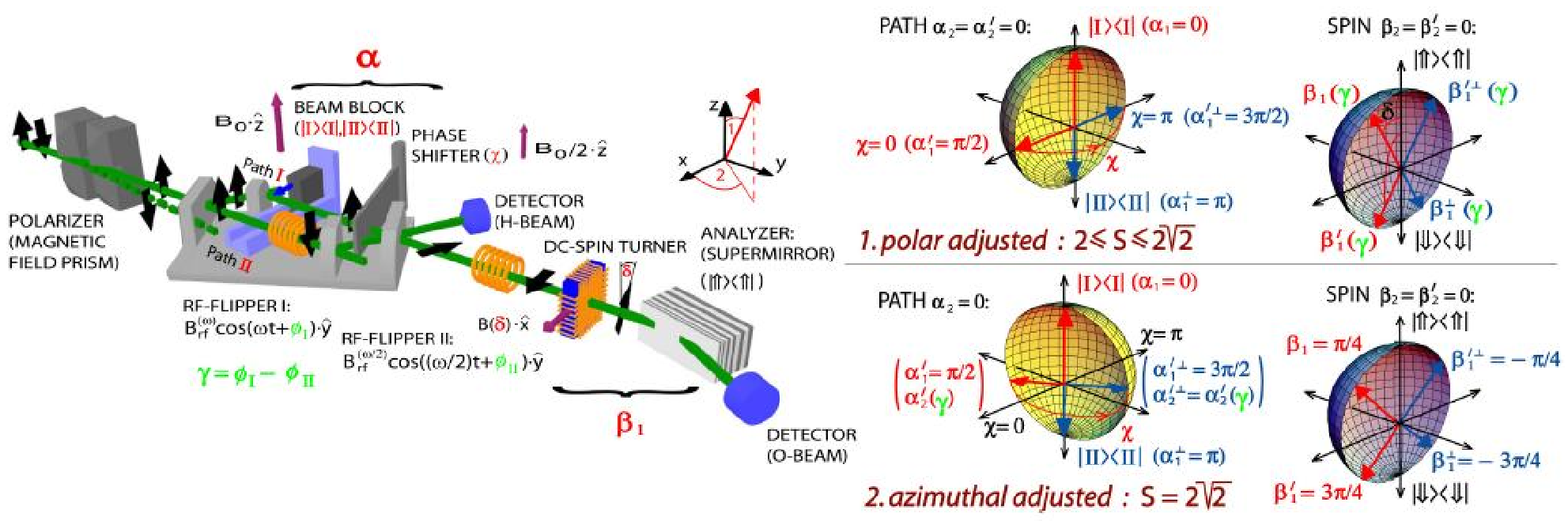}} \caption {The experimental apparatus for joint measurement of spinor and path degrees of freedom
with respect to the geometric phase. The incident neutron beam is polarized by a magnetic field prism. The spin state acquires a geometric phase
$\gamma$ during the interaction with the two rf-fields and is flipped twice. The beam block is required for measurements solely in one path
($\pm\,\hat {\mathbf z}$ direction of the path measurement). Finally, the spin is rotated by an angle $\delta$ (in the $\hat{\mathbf x},
\hat{\mathbf z}$ plane), by a dc-spin turner, for a polarization analysis and count rate detection. The Bloch-sphere description includes the
measurement settings of $\boldsymbol\alpha$ and $\boldsymbol\beta(\delta)$, determining the projection operators, used for joint measurement of
spin and path. $\boldsymbol\alpha$ is tuned by a combination of the phase shifter ($\chi$) and the beam block, and $\boldsymbol\beta$ is
adjusted by the angle $\delta$.}\label{fig:setup}
\end{figure*}
%
%

\subsection{Bell-like Inequalities}

Next, a Bell-like inequality in CHSH-formalism \cite{Clauser69} is introduced, consisting of four expectation values with the associated
directions $\boldsymbol\alpha$, $\boldsymbol\alpha'$ and $\boldsymbol\beta$, $\boldsymbol\beta'$ for joint measurements of spin and path,
respectively
\begin{eqnarray}\label{eq:S-functionOrg}
\begin{split}
& S(\boldsymbol\alpha,\boldsymbol\alpha',\boldsymbol\beta,\boldsymbol\beta',\gamma)=\big\vert
E(\boldsymbol\alpha,\boldsymbol\beta)-E(\boldsymbol\alpha,\boldsymbol\beta')
 + E(\boldsymbol\alpha',\boldsymbol\beta)\\&+E(\boldsymbol\alpha',\boldsymbol\beta')\big\vert
 \\&=\Bigl\vert-\sin\alpha_1\bigl(\cos(\alpha_2-\beta_2-\gamma)\sin\beta_1
 \\&-\cos(\alpha_2-\beta'_2\gamma)\sin\beta'_1\bigr)
 -\cos\alpha_1\bigl(\cos\beta_1-\cos\beta'_1\bigr)
  \\&-\sin\alpha'_1\bigl(\cos(\alpha'_2-\beta_2-\gamma)\sin\beta_1
 +\cos(\alpha'_2-\beta'_2-\gamma)\sin\beta'_1\bigr)
 \\&-\cos\alpha'_1\bigl(\cos\beta_1+\cos\beta'_1\bigr)\Bigr\vert.
\end{split}
\end{eqnarray}

The boundary of Eq.(\ref{eq:S-functionOrg}) is given by the value 2 for any NCHVT \cite{Basu01}. Without loss of generality one angle can be
eliminated by setting, e.g., $\boldsymbol\alpha=0$ ($\alpha_1=\alpha_2=0$), which gives
\begin{eqnarray}\label{S-function1}
 \begin{split}
    S(\boldsymbol\alpha',\boldsymbol\beta,\boldsymbol\beta',\gamma)&=\Bigl\lvert -\sin\alpha'_1\Bigl(\cos(\alpha'_2-\beta_2-\gamma)\sin\beta_1\\
    &+\cos(\alpha'_2-\beta'_2-\gamma)\sin\beta'_1\Bigr)-\cos\alpha'_1\\
    &\times(\cos\beta_1+\cos\beta'_1)  -\cos\beta_1+\cos\beta'_1\Bigr\rvert.
\end{split}
\end{eqnarray}
Keeping the polar angles $\alpha'_1$, $\beta_1$ and $\beta'_1$ constant at the usual Bell angles $\alpha'_1=\frac{\pi}{2}$,
$\beta_1=\frac{\pi}{4}$, $\beta'_1=\frac{3\pi}{4}$ (and azimuthal parts fixed at $\alpha'_2=\beta_2=\beta'_2=0$) reduces $S$ to
\begin{eqnarray}
    S(\gamma)= \big\lvert-\sqrt{2}-\sqrt{2}\cos\gamma\big\rvert,
\end{eqnarray}
where the familiar maximum value of $2\sqrt{2}$ is reached for $\gamma=0$. For $\gamma=\pi$ the value of $S$ approaches zero.

\subsubsection{\label{sec:level3}Polar Angle Adjustment}

Here we consider the case when the azimuthal angles are kept constant, e.g., $\alpha'_2=\beta_2=\beta'_2=0$ ($\alpha_2=0)$, denoted as
\begin{eqnarray}\label{eq:S-functionAzFixed}
 \begin{split}
S(\alpha'_1,\beta_1,\beta'_1,\gamma)&=\Bigl\lvert -\sin\alpha'_1\Bigl(\cos\gamma\sin\beta_1+\cos\gamma\sin\beta'_1\Bigr)\\
&-\cos\alpha'_1(\cos\beta_1+\cos\beta'_1) \\
&  -\cos\beta_1+\cos\beta'_1\Bigr\rvert.
\end{split}
\end{eqnarray}
The polar Bell angles $\beta_1$, $\beta'_1$ and $\alpha'_1$ ($\alpha_1=0$), yielding a maximum $S$-value, can be determined, with respect to the
geometric phase $\gamma$, by calculating the partial derivatives (the extremum condition) of $S$ in Eq.(\ref{eq:S-functionAzFixed}):
\begin{eqnarray}
    \frac{\partial S}{\partial\beta_1}  &=&
    \sin\beta_1+\cos\alpha'_1\sin\beta_1-\cos\gamma\sin\alpha'_1\cos\beta_1=0\nonumber\\
    \frac{\partial S}{\partial\beta'_1} &=&
    -\sin\beta'_1+\cos\alpha'_1\sin\beta'_1-\cos\gamma\sin\alpha'_1\cos\beta'_1=0\nonumber\\
    \frac{\partial S}{\partial\alpha'_1} &=&
    \sin\alpha'_1(\cos\beta_1+\cos\beta'_1)\nonumber\\&-&\cos\gamma\cos\alpha'_1
    (\sin\beta_1+\sin\beta'_1)=0.
\end{eqnarray}
The solutions are given by
\begin{subequations}
\begin{eqnarray}\label{eq:Bell-angles}
\beta_1 &=& \arctan\big(\cos\gamma\big)\label{eq:Cond1}\\
\beta'_1 &=& \pi - \beta_1\label{eq:Cond2}\\
\alpha'_1 &=& \frac{\pi}{2}\,.
\end{eqnarray}
\end{subequations}

With these angles the maximal $S$ decreases for $\gamma : 0\rightarrow\frac{\pi}{2}$ and touches at $\gamma = \frac{\pi}{2}$ even the limit of
the CHSH inequality $S=2$.

\subsubsection{\label{sec:level3}Azimuthal Angle Adjustment}

Next we discuss the situation where the standard maximal value $S=2\sqrt{2}$ can be achieved by keeping the polar angles $\alpha'_1$, $\beta_1$
and $\beta'_1$ constant at the Bell angles $\alpha'_1=\frac{\pi}{2}$, $\beta_1=\frac{\pi}{4}$, $\beta'_1=\frac{3\pi}{4}$, ($\alpha_1=0$), while
the azimuthal parts, $\alpha'_2$, $\beta_2$ and $\beta'_2$ ($\alpha_2=0$), are varied. The corresponding S function is denoted as
\begin{eqnarray}
 \begin{split}
    S(\alpha'_2,\beta_2,\beta'_2,\gamma)
    &= \Bigl\lvert-\sqrt{2}-\frac{\sqrt{2}}{2}\Bigl(\cos(\alpha'_2-\beta_2-\gamma)\\&+\cos(\alpha'_2-\beta'_2-\gamma)\Bigr)\Bigr\rvert\;.
\end{split}
\end{eqnarray}

The maximum value $2\sqrt{2}$ is reached for
\begin{subequations}
\begin{eqnarray}
\beta_2&=&\beta'_2\textrm{, and}\label{eq:Cond3}\\
\alpha'_2-\beta'_2&=&\gamma\, (\textrm{mod}\,\pi)\label{eq:Cond4}.
\end{eqnarray}
\end{subequations}
 For convenience $\beta_2=0$ is
chosen.

The conditions expressed in Eq.(\ref{eq:Cond1}), Eq.(\ref{eq:Cond2}) and Eq.(\ref{eq:Cond4}) (see Bloch spheres in Fig.\,\ref{fig:setup}) are
experimentally realized using the spin turner device and the neutron interferometer depicted in Fig.\,\ref{fig:setup}.

\section{\label{sec:experiment}Description of the experiment}

\subsection{\label{sec:preparation}State Preparation}

The preparation of entanglement between spatial and the spinor degrees of freedom is achieved by a beam splitter and a subsequent spin flip
process in one sub beam: Behind the beam splitter (first plate of the IFM) the neutron\char39{}s wave function is found in a coherent
superposition of $\ket{\textrm{I}}$ and $\ket{\textrm{II}}$, and only the spin in $\ket{\textrm{II}}$ is flipped by the first rf-flipper within
the interferometer (see Fig.\,\ref{fig:setup}).

%
\begin{figure} [t]
\scalebox{0.2}{\includegraphics{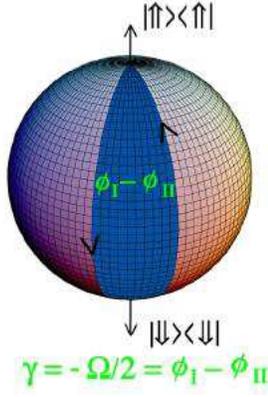}} \caption{\label{fig:geoPhase} Schematic representation of the spinor evolution, within the two
successive rf flippers, on the Bloch sphere. The geometric phase $\gamma$ is given by minus half of the solid angle $\Omega$, traced out by the
state vector, depending on the phases $\phi_{\textrm{I}}$ and $\phi_{\textrm{II}}$ of the oscillating magnetic fields in the rf flippers.}
\end{figure}
%
%
%
The entangled state which emerges from a coherent superposition of $\ket{\textrm{I}}$ and $\ket{\textrm{II}}$ is expressed as
$\ket{\Psi_{\textrm{Bell}}}=\frac{1}{\sqrt{2}}\Big(\ket{\textrm{I}}\otimes\ket{\Uparrow}+\ket{\textrm{II}}\otimes
e^{i\phi_{\textrm{I}}}\ket{\Downarrow}\Big)$, where the state vector of the neutron acquires a phase $\phi_{\textrm{I}}$ during the interaction
with the oscillating field, given by $B^{(1)}=B_{\textrm{rf}}^{(\omega)} \cos(\omega t +\phi_{\textrm{I}})\cdot\hat{\mathbf y}$ (for a more
detailed description of the generation of $\ket{\Psi_{\textrm{Bell}}}$ see \cite{Sponar07}).

\subsection{\label{sec:geo}Manipulation of Geometric and Dynamical Phases}

Within the rf-flipper placed inside the interferometer (path II), the neutron spin traces out a semi-great circle from $\ket{\Uparrow}$ to
$\ket{\Downarrow}$ on the Bloch sphere and returns to its initial state $\ket{\Uparrow}$ when passing the second rf-flipper (see
Fig.\,\ref{fig:geoPhase}). The two semi-great circles enclose an angle $\phi_{\textrm{I}}-\phi_{\textrm{II}}$ (orange slice), and hence a solid
angle $\Omega=2(\phi_{\textrm{I}}-\phi_{\textrm{II}})$. The solid angle $\Omega$ yields a pure geometric phase $\gamma=-\Omega/2$ as in
\cite{Badurek2000GeoPhaseSep,Allman97}. Since we set $\phi_{\textrm{II}}=0$ the geometric phase is given by $\gamma=\phi_{\textrm{I}}$ and the
state is represented by
\begin{equation}\label{eq:stateExp}
\ket{\Psi_{\textrm{Exp}}(\gamma)}=\frac{1}{\sqrt{2}}\Big(\ket{\textrm{I}}\otimes\ket{\Uparrow}+\ket{\textrm{II}}\otimes
e^{i\gamma}\ket{\Downarrow}\Big).
\end{equation}
In our experiment the $\ket{\Uparrow}$ eigenstate (in path I and II) acquires dynamical phase as it precesses about the magnetic guide field in
$+\,\hat{\mathbf z}$-direction. After a spin-flip (only in path II) the $\ket{\Downarrow}$ eigenstate still gains another dynamical phase but of
opposite sign compared to the situation before the spin-flip. These phases manifest in a dynamical phase offset, which remains constant during
the complete measurement procedure.

\subsection{\label{sec:jointMeasure}Joint Measurements}

Experimentally, the probabilities of joint (projective) measurements are proportional to the following count rates
\begin{widetext}
\begin{eqnarray}
 \begin{split}
    &N_{++}(\boldsymbol\alpha,\boldsymbol\beta)=N_{++}\big(\boldsymbol\alpha,(\beta_1,0)\big)\propto\langle
    \Psi_{\textrm{Exp}}(\gamma)\rvert \hat P_+^{\textrm{p}}(\boldsymbol\alpha)\otimes
    \hat P_+^{\textrm{s}}(\beta_1,0)\lvert\Psi_{\textrm{Exp}}(\gamma)\rangle\\
    &N_{+-}(\boldsymbol\alpha,\boldsymbol\beta)
    =N_{++}\big(\boldsymbol\alpha,(\beta_1+\pi,0)\big)\equiv N_{++}\big(\boldsymbol\alpha,(\beta^\bot_1,0)\big)\propto
    \langle\Psi_{\textrm{Exp}}(\gamma)\rvert \hat P_+^{\textrm{p}}(\boldsymbol\alpha)\otimes
    \hat P_+^{\textrm{s}}(\beta^\bot_1,0)\lvert\Psi_{\textrm{Exp}}(\gamma)\rangle\\
    &N_{-+}(\boldsymbol\alpha,\boldsymbol\beta)=
    N_{++}\big((\alpha_1+\pi,\alpha_2),(\beta_1,0)\big)\equiv N_{++}\big((\alpha^\bot_1,\alpha_2),(\beta_1,0)\big)\propto\langle\Psi_{\textrm{Exp}}(\gamma)\rvert \hat P_+^{\textrm{p}}
    (\alpha^\bot_1,\alpha_2)\otimes
    \hat P_+^{\textrm{s}}(\beta_1,0)\lvert\Psi_{\textrm{Exp}}(\gamma)\rangle\\
    &N_{--}(\boldsymbol\alpha,\boldsymbol\beta)
    =N_{++}\big((\alpha^\bot_1,\alpha_2),(\beta^\bot_1,0)\big)
    \propto
    \langle\Psi_{\textrm{Exp}}(\gamma)\rvert \hat P_+^{\textrm{p}}(\alpha^\bot_1,\alpha_2)\otimes
    \hat P_+^{\textrm{s}}(\beta^\bot_1,0)\lvert\Psi_{\textrm{Exp}}(\gamma)\rangle.
\end{split}
\end{eqnarray}

\end{widetext} The expectation value of a joint measurement of $A^{\textrm{p}}(\boldsymbol\alpha)$ and $B^{\textrm{s}}(\boldsymbol\beta)$
\begin{equation}
    E(\boldsymbol\alpha,\boldsymbol\beta)=\langle\Psi(\gamma)\rvert A^{\textrm{p}}(\boldsymbol\alpha)\otimes
    B^{\textrm{s}}(\boldsymbol\beta)\lvert\Psi(\gamma)\rangle
\end{equation}
is experimentally determined from the count rates
\begin{eqnarray}\label{expectation_value_exp}
 \begin{split}
   &E(\boldsymbol\alpha,\boldsymbol\beta)=\\
&\frac{N_{++}(\boldsymbol\alpha,\boldsymbol\beta)-N_{+-}(\boldsymbol\alpha,\boldsymbol\beta)-N_{-+}(\boldsymbol\alpha,\boldsymbol\beta)
+N_{--}(\boldsymbol\alpha,\boldsymbol\beta)}
    {N_{++}(\boldsymbol\alpha,\boldsymbol\beta)+N_{+-}(\boldsymbol\alpha,\boldsymbol\beta)+N_{-+}(\boldsymbol\alpha,\boldsymbol\beta)
    +N_{--}(\boldsymbol\alpha,\boldsymbol\beta)}\;.
 \end{split}
\end{eqnarray}
With these expectation values S is defined by
\begin{equation}\label{eq:S_Value}
S=E(\boldsymbol\alpha,\boldsymbol\beta)-E(\boldsymbol\alpha,\boldsymbol\beta')+E(\boldsymbol\alpha',\boldsymbol\beta)+E(\boldsymbol\alpha',\boldsymbol\beta').
\end{equation}

\subsection{\label{sec:jointMeasure}Experimental setup}

The experiment was carried out at the neutron interferometer instrument S18 at the high-flux reactor of the Institute Laue-Langevin (ILL) in
Grenoble, France. A sketch of the setup is depicted in Fig.\,\ref{fig:setup}. A monochromatic beam, with mean wavelength $\lambda_0=1.91 \mbox{
\AA} (\Delta\lambda/\lambda_0\sim0.02$) and 5x5 mm$^2$ beam cross-section, is polarized by a bi-refringent magnetic field prism in $\hat{\mathbf
z}$-direction \cite{Badurek00FieldPrism}. Due to the angular separation at the deflection, the interferometer is adjusted so that only the
spin-up component fulfills the Bragg condition at the first interferometer plate (beam splitter).

As in our previous experiment \cite{Sponar07}, the spin in path $\ket{\textrm{II}}$ is flipped by a rf-flipper, which requires two magnetic
fields: A static field $B_0\cdot\hat{\mathbf z}$ with $B_0=\hbar \omega/(2\lvert\mu\lvert)$ and a perpendicular oscillating field
$B^{(1)}=B_{\textrm{rf}}^{(\omega)} \cos(\omega t +\phi_{\textrm{I}})\cdot\hat{\mathbf y}$ with amplitude
$B^{(\omega)}_{\textrm{rf}}=\pi\hbar/(2\tau\lvert\mu\lvert$), where $\mu$ is the magnetic moment of the neutron and $\tau$ denotes the time the
neutron is exposed to the rf-field. The oscillating field is produced by a water-cooled rf-coil with a length of 2\,cm, operating at a frequency
of $\omega/2\pi=58$\,kHz. The static field is provided by a uniform magnetic guide field $B_0\sim 2$\,mT, produced by a pair of water-cooled
Helmholtz coils.

The two sub-beams are recombined at the third crystal plate where $\ket{\textrm{I}}$ and $\ket{\textrm{II}}$ only differ by an adjustable phase
factor $e^{i\chi}$ (path phase $\chi$ is given by $\chi=N b_c\lambda D$, with the thickness of the phase shifter plate $D$, the neutron
wavelength $\lambda$, the coherent scattering length $b_c$ and the particle density $N $ in the phase shifter plate). By rotating the plate,
$\chi$ can be varied systematically. This yields the well known intensity oscillations of the two beams emerging behind the interferometer.

The O-beam passes the second rf-flipper, operating at $\omega/2\pi=29$\,kHz, which is half the frequency of the first rf-flipper. The
oscillating field is denoted as $B^{(\omega/2)}_{\textrm{rf}} \cos\big((\omega/2) t +\phi_{\textrm{II}}\big)\cdot\hat{\mathbf y}$, and the
strength of the guide field was tuned to about $1\,$mT in order to satisfy the frequency resonance condition. This flipper compensates the
energy difference between the two spin components, by absorption and emission of photons of energy $E=\hbar\omega/2$ (see \cite{Sponar07}).

Finally, the spin is rotated by an angle $\delta$ (in the $\hat{\mathbf x}, \hat{\mathbf z}$ plane) with a static field spin-turner, and
analyzed due to the spin dependent reflection within a Co-Ti multi-layer supermirror along the $\hat{\mathbf z}$-direction. With this
arrangement consisting of a dc-spin turner and a supermirror the spin can be analyzed along arbitrary directions in the $\hat{\mathbf x},
\hat{\mathbf z}$ plane, determined by $\delta$, which is measured from the $ \hat{\mathbf z}$ axis (see Fig.\,\ref{fig:setup}, and later front
panel of Fig.\,\ref{fig:OsciPolAdjust} for intensity modulations due to $\chi$-scans).

\subsection{\label{sec:jointMeasure}Experimental Strategy}

\subsubsection{\label{sec:level3}Polar Angle Adjustment}
%
\begin{figure}[b]
\scalebox{0.43}{\includegraphics{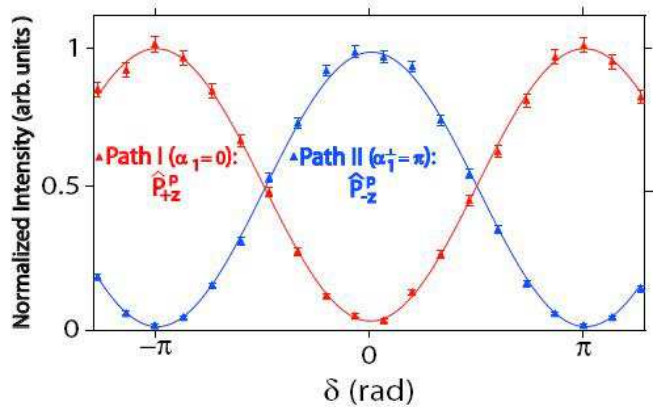}} \caption{\label{fig:OsciPolandAzimuthalAdjust} Typical intensity modulations obtained by inserting a
beam block, being the projections on the $\pm\,\hat{\mathbf z}$ direction of the path measurement, denoted as $\hat
P^{\textrm{p}}_{+z}:({\alpha_1=0, \alpha_2=0})$ and $\hat P^{\textrm{p}}_{-z}:({\alpha_1^\bot=\pi, \alpha_2=0})$. The oscillations remain the
same when altering the geometric phase.}
\end{figure}
%
%

\begin{figure*}[t]
\scalebox{0.89}{\includegraphics{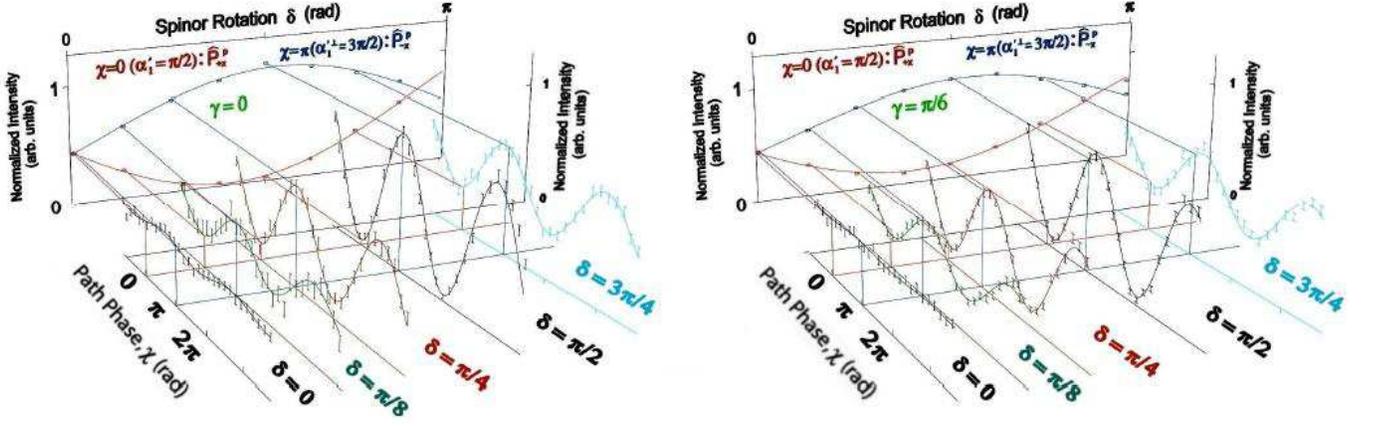}} \caption{\label{fig:OsciPolAdjust} Typical interference patterns of the O-beam ($\alpha'_1=\pi/2$)
for $\delta=0,\pi/8,\pi/4,\pi/2,3\pi/4$, being the direction of the spin analysis, and geometric phase $\gamma=0$ (left) and $\gamma=\pi/6$
(right side). Intensities at the path phase $\chi= 0$ and $\chi=\pi$ are extracted from least square fits of the oscillations. The resulting
curves (rear panel) represent the projections to the $\pm\,\hat{\mathbf x}$ direction of the path subspace, denoted as $\hat
P^{\textrm{p}}_{+x}:({\alpha'_1=\pi/2, \alpha'_2=0})$ and $\hat P^{\textrm{p}}_{-x}:({\alpha'^{\bot}_1=3\pi/2, \alpha'_2=0})$. The shift of the
oscillations (see for instance $\delta=\pi/2$), due to the geometric phase $\gamma$, yields a lower contrast of the curves $\hat
P^{\textrm{p}}_{+x}$ and $\hat P^{\textrm{p}}_{-x}$.}
\end{figure*}

Projective measurements are performed on parallel planes defined by $\alpha_2=\alpha'_2=\beta_2=\beta'_2=0$ (see Fig.\,\ref{fig:setup}). For the
path measurement the directions are given by $\boldsymbol\alpha:\alpha_1=0, \alpha_2=0$ (Fig.\,\ref{fig:OsciPolandAzimuthalAdjust}), and
$\boldsymbol\alpha': \alpha'_1=\pi/2,\,\alpha'_2=0$ (Fig.\,\ref{fig:OsciPolAdjust}).
%
%
%
%
%

The angle $\boldsymbol\alpha$, which corresponds to $+\,\hat{\mathbf z}$ (and $-\,\hat{\mathbf z}$ for
$\alpha_1^\bot=\alpha_1+\pi=\pi,\alpha_2=0$) is achieved by the use of a beam block which is inserted to stop beam II (I) in order to measure
along $+\,\hat{\mathbf z}$ (and $-\,\hat{\mathbf z}$). The corresponding operators are given by
\begin{eqnarray}
\hat P^{\textrm{p}}_{+z}({\alpha_1=0, \alpha_2=0})&=&\ketbra{\textrm{I}}{\textrm{I}}\nonumber\\
\hat P^{\textrm{p}}_{-z}({\alpha_1^\bot=\pi, \alpha_2=0})&=&\ketbra{\textrm{II}}{\textrm{II}},
\end{eqnarray}
The results of the projective measurement are plotted versus different angles $\delta$ of the spin analysis, which is depicted in
Fig.\,\ref{fig:OsciPolandAzimuthalAdjust}. Complementary oscillations were obtained due to the spin flip in path $\ket{\textrm{II}}$. These
curves are insensitive to the geometric phase $\gamma$, due to the lack of superposition with a referential sub-beam.

The angle $\boldsymbol\alpha'$ is set by a superposition of equal portions of $\ket{\textrm{I}}$ and $\ket{\textrm{II}}$, represented on the
equator of the Bloch sphere. The interferograms are achieved by a rotation of the phase shifter plate, associated with a variation of the path
phase $\chi$, repeated at different values of the spin analysis direction $\delta$. The projective measurement for $\alpha'_1=\pi/2,\alpha'_2=0$
corresponds to a phase shifter position of $\chi$=~0 (and $\alpha'_1$$^\bot$=~$\alpha'_1+\pi=3\pi/2,\alpha'_2=~0$ to $\chi=\pi$). Projection
operators read as
\begin{eqnarray}
 \begin{split}
&\hat P^{\textrm{p}}_{+x}({\alpha'_1=\frac{\pi}{2}, \alpha'_2=0})=\frac{1}{2}\Big(\big(\ket{\textrm{I}}+\ket{\textrm{II}}\big)\big(\bra{\textrm{I}}+\bra{\textrm{II}}\big)\Big)\\
&\hat P^{\textrm{p}}_{-x}({\alpha'^{\bot}_1=\frac{3\pi}{2}, \alpha'_2=0})=
\frac{1}{2}\Big(\big(\ket{\textrm{I}}-\ket{\textrm{II}}\big)\big(\bra{\textrm{I}}-\bra{\textrm{II}}\big)\Big).
 \end{split}
\end{eqnarray}

The interferogram obtained for $\gamma=0$ and $\delta=\pi/2$, in Fig.\,\ref{fig:OsciPolAdjust}, is utilized to determine the zero point of the
path phase $\chi$, which defines the $+\,\hat{\mathbf x}$\,-\,direction ($\alpha'_1=\pi/2,\alpha_2'=0$) for the path measurement.

In order to obtain phase shifter scans of higher accuracy, scans over two periods were recorded (see Fig.\,\ref{fig:OsciPolAdjust}) and the
values for $\chi=0$ and $\pi$ are extracted from the data by least square fits. These extracted points, marking the $\pm\,\hat{\mathbf
x}$-direction of the path measurement, are plotted versus different angles of $\delta$, as shown in Fig.\,\ref{fig:OsciPolAdjust}, rear diagram.

All phase shifter scans were repeated for different angles $\delta$ for the spin analysis from $\delta$=0 to $\delta=\pi$ in steps of $\pi/8$,
and for several geometric phases $\gamma$ (steps of $\pi/6$, and beginning form $\gamma=\pi$ steps of $\pi/4$ ), depicted in the rear panel of
Fig.\,\ref{fig:OsciPolAdjust} for five selected settings of $\delta$ ($\delta=0,\pi/8,\pi/4,\pi/2,3\pi/4$) and two geometric phases
($\gamma=0,\pi/6$).

\subsubsection{\label{sec:level3}Azimuthal Angle Adjustment}

%
%
\begin{figure}[b]
\scalebox{0.42}{\includegraphics{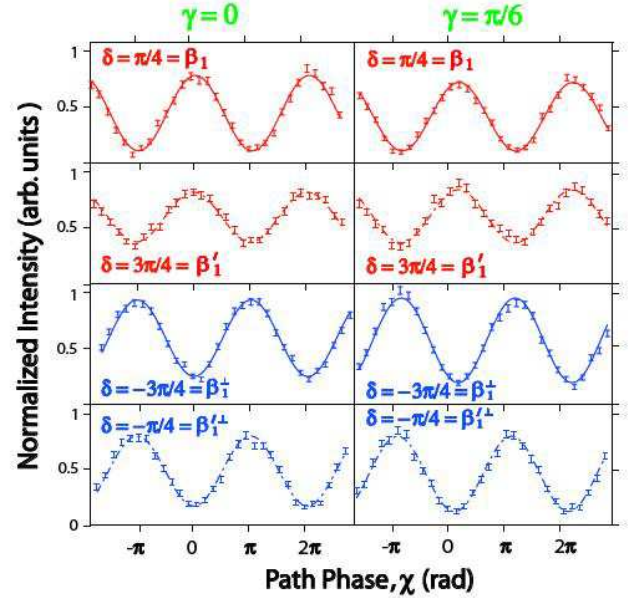}} \caption{\label{fig:OsciAzimuthAdjust} Typical interference patterns of the O-beam
($\alpha'_1=\pi/2$) for $\delta=\pi/4=\beta_1$, $\delta=3\pi/4=\beta'_1$, $\delta=-3\pi/4=\beta_1^\bot$ and $\delta=-\pi/4=\beta'_1$$^\bot$
($\beta_2=\beta'_2=0$) and geometric phase $\gamma=0$~(left) and $\gamma=\pi/6$~(right). Phase shifter scans $\chi$ are performed for a
forthcoming determination of $\alpha'_2$.}
\end{figure}
%
%
%
%
\begin{figure}[b]
\scalebox{0.42}{\includegraphics{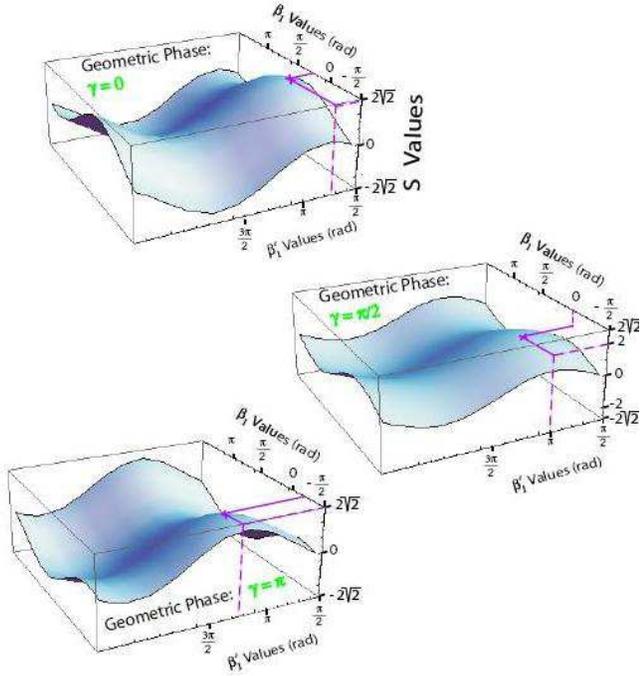}} \caption{\label{fig:SValuePolAdjust}$S$-values for different settings of geometric phases $\gamma$,
derived from the least square fits of the projective measurements along $\pm\,\hat{\mathbf z}$ (beam block
Fig.\,\ref{fig:OsciPolandAzimuthalAdjust}) and $\pm\,\hat{\mathbf x}$ (varying $\chi$ Fig.\,\ref{fig:OsciPolAdjust}) direction for the path
measurement ($\alpha_1=0$, $\alpha'_1=\pi/2$ and $\alpha_2=\alpha'_2=0$) using Eq.(\ref{eq:S-functionAzFixed}). $\beta_1$ and $\beta'_1$
represent the direction of the spin analysis, which are changed systematically by a variation of $\delta$ while $\beta_2$ and $\beta'_2$ remain
constant at the value zero (\emph{polar adjustment}). The position of the maximum is determined numerically for different settings the geometric
phases $\gamma$ (here for example $\gamma=0$ where $S_{\text{MAX}}=2\sqrt{2}$, $\gamma=\pi/2$ with $S_{\text{MAX}}=2$, and $\gamma=\pi$, with
$S_{\text{MAX}}=2\sqrt{2}$). The $\beta_1$ and $\beta'_1$ values result as predicted in Eqs.(\ref{eq:Cond1}) and (\ref{eq:Cond2}) for
$S_{\text{MAX}}$.}
\end{figure}
%
%

Here the Bell angles (polar angles) remain fixed at the usual values and are set at $\delta$ for the projective spin measurement, and by the
beam block (and fixed phase shifter positions) for the path measurement. The angle between the measurement planes is adjusted by one azimuthal
angle ($\alpha'_2$), which is deduced by phase shifter ($\chi$) scans.

For the spin measurement the directions are fixed and given by given by $\boldsymbol\beta$: $\beta_1=\pi/4$, $\beta_2=0$ and
$\boldsymbol\beta'$: $\beta'_1=3\pi/4$, $\beta'_2=0$ (together with  $\beta^\bot_1=-3\pi/4$, $\beta'^{\bot}_1 =-\pi/4$, see
Fig.\,\ref{fig:setup} for Bloch description and Fig.\,\ref{fig:OsciAzimuthAdjust} for measured interference patterns). For the projective path
measurement the fixed directions read as $\alpha_1=0$ ($\alpha_1^\bot=\pi$, see Fig.\,\ref{fig:OsciPolandAzimuthalAdjust} for measurements with
beam block), and $\alpha'_1=\pi/2$ ($\alpha'^\bot_1=3\pi/2$). Phase shifter ($\chi$) scans are performed in order to determine $\alpha'_2$,
which is depicted in Fig.\,\ref{fig:OsciAzimuthAdjust} for two values of the geometric phase: $\gamma=0$ and $\gamma=\pi/6$. One can see a shift
of the oscillations due to the geometric phase $\gamma$.

\subsection{\label{sec:analysis}Data Analysis and Experimental Results}

\subsubsection{\label{sec:level3}Polar Angle Adjustment}

Using least square fits from the \emph{polar angle adjustment} measurement curves in Fig.\,\ref{fig:OsciPolAdjust} and
Fig.\,\ref{fig:OsciPolandAzimuthalAdjust}, together with Eq.(\ref{eq:S_Value}) the $S$-value is calculated as a function of the parameters
$\beta_1$ and $\beta_1'$ which is plotted in Fig.\,\ref{fig:SValuePolAdjust} for $\gamma=0$, $\gamma=\pi/2$ and $\gamma=\pi$ ( $\gamma=0$ and
$\gamma=\pi$ are chosen since the fringe displacement is maximal for these two settings and $\gamma=\pi/2$ illustrates the increase of $S$ to a
value of 2). The local maximum of the surface is determined numerically. The settings for $\beta_1$ and $\beta'_1$, yielding a maximal
$S$-value, are compared with the predicted values for $\beta_1$ and $\beta'_1$ from Eqs.(\ref{eq:Cond1}) and (\ref{eq:Cond2}), respectively.

\begin{figure}[t]
\scalebox{0.45}{\includegraphics{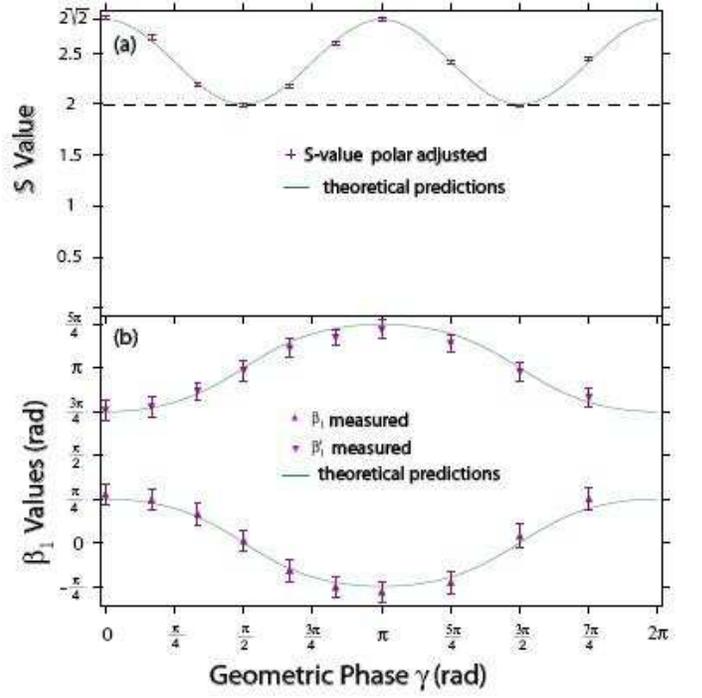}} \caption{\label{fig:ResultsPolAdjust}(a) \emph{Polar adjusted} $S$-values versus geometric phase
$\gamma$ with adapted Bell angles ($\beta_1$ and $\beta'_1$) according to the geometric phase $\gamma$. (b) the corresponding modified Bell
angles are plotted versus the geometric phase $\gamma$.}
\end{figure}

The resulting $S$ values, derived by using the adjusted Bell angles $\beta_1$ and $\beta'_1$, are plotted in
Fig.\,\ref{fig:ResultsPolAdjust}\,(a) versus the geometric phase $\gamma$. The theoretical predictions from Eq.(\ref{eq:S-functionAzFixed})
depicted as solid (color: green) line are evidently reproduced. The maximal $S$ decreases from $\gamma$=0 to $\gamma=\pi/2$ where the boundary
of the CHSH inequality $S=2$ is reached, followed by an increase to the familiar value $S=2\sqrt{2}$.

%
%
%

In Fig.\,\ref{fig:ResultsPolAdjust}\,(b) the deduced $\beta_1$ and $\beta'_1$ values are plotted versus the geometric phase $\gamma$. $\beta_1$
and $\beta'_1$ follow the theoretical behavior (solid line - color: green line), predicted by Eqs.(\ref{eq:Cond1}) and (\ref{eq:Cond2}). One can
see a peak for $\beta_1$ (and a dip for $\beta'_1$) at $\gamma=\pi$.

\subsubsection{\label{sec:level3}Azimuthal Angle Adjustment}

\begin{figure}[b]
\scalebox{0.43}{\includegraphics{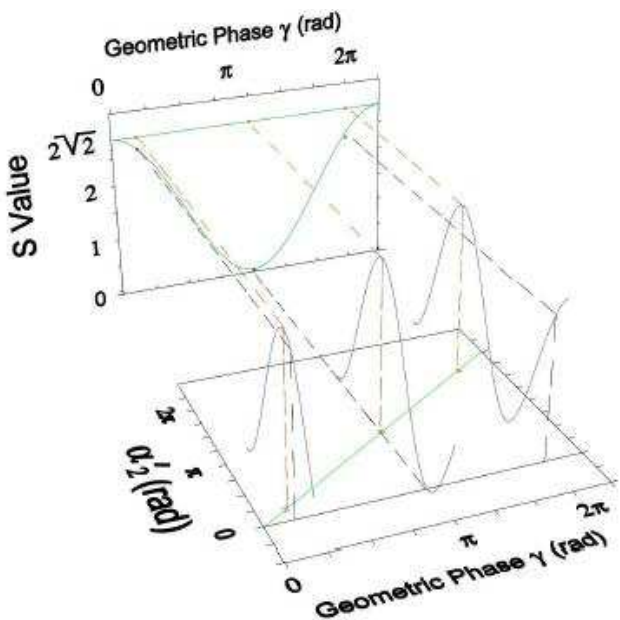}} \caption{\label{fig:SValueAzimuthAdjust} $S$-values derived from least square fits of the projective
spin and path measurements for $\beta_1=\pi/4$, $\beta'_1=3\pi/4$, $\beta_1^\bot=-3\pi/4$, $\beta'_1$$^\bot=-\pi/4$ and
$\alpha'_1=\pi/2,~\alpha'^{\bot}_1=3\pi/2$ (see Fig.\,\ref{fig:OsciAzimuthAdjust}), and $\alpha_1=0$, $\alpha^\bot_1=\pi$ (see
Fig.\,\ref{fig:OsciPolandAzimuthalAdjust}), versus geometric phase $\gamma$. The maximum S-value of $2\sqrt{2}$ is always found for
$\alpha'_2=\gamma$ as predicted in Eq.(\ref{eq:Cond3}) (\emph{azimuthal adjustment}). If no corrections are applied to the Bell angles
($\alpha'_2=0$) S approaches zero at $\gamma=\pi$. }
\end{figure}

\begin{figure}[t]
\scalebox{0.42}{\includegraphics{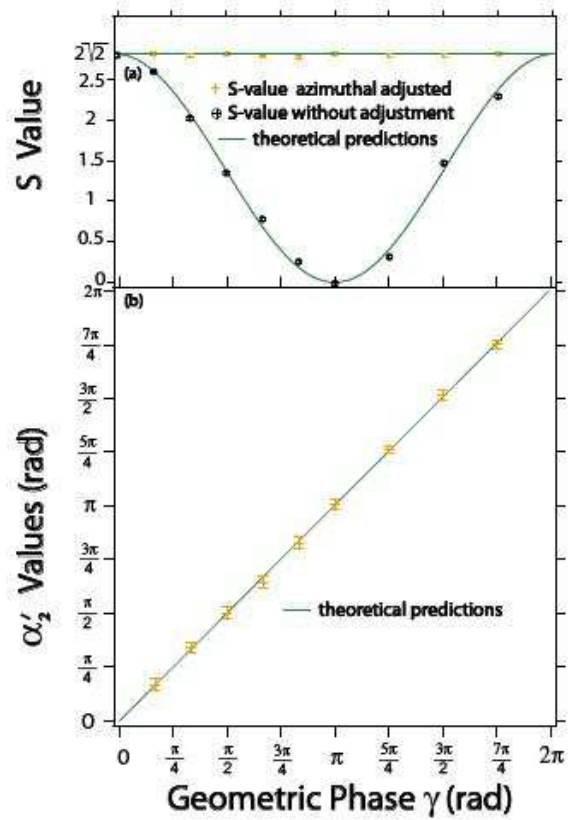}} \caption{\label{fig:ResultsAzimuthAdjust} (a) \emph{Azimuthal adjusted} $S$-values versus geometric
phase $\gamma$ with balanced Bell angle ($\alpha'_2$) according to the geometric phase $\gamma$, and without corrections. (b) the corresponding
modified Bell angle is plotted versus the geometric phase $\gamma$.}
\end{figure}

In Fig.\,\ref{fig:SValueAzimuthAdjust} we depict selected $S$ values calculated from least square fits of the \emph{azimuthal angle adjustment}
measurements, where $\beta_1=\pi/4$, $\beta'_1=3\pi/4$, $\beta_1^\bot=5\pi/4$, $\beta'_1$$^\bot=-\pi/4$ and $\alpha'_1=\pi/2$ (see
Fig.\,\ref{fig:OsciAzimuthAdjust}) and $\alpha_1=0$, $\alpha^\bot_1=\pi$ (Fig.\,\ref{fig:OsciPolandAzimuthalAdjust}) versus geometric phase
$\gamma$. A simple shift of the oscillation of the S-value is observed due to the geometric phase (see Fig.\,\ref{fig:SValueAzimuthAdjust} front
panel). The maximum $S$-value of $2\sqrt{2}$ is always found for $\alpha'_2=\gamma$, which is indicated in the rear panel of
Fig.\,\ref{fig:SValueAzimuthAdjust}. The complete measurement set of $S$-values versus the geometric phase $\gamma$ is plotted in
Fig.\,\ref{fig:ResultsAzimuthAdjust}\,(a)-$S$-\emph{value azimuthal adjusted}. If no adjustment is applied to $\alpha_2$, which means
$\alpha'_2$ is always kept constant at $\alpha_2=0$, $S$ approaches zero at $\gamma=\pi$ and returns to the maximum value $2\sqrt2$ at
$\gamma=2\pi$ (see Fig.\,\ref{fig:ResultsAzimuthAdjust}\,(a)-$S$-\emph{value without adjustment}).

Figure~\ref{fig:ResultsAzimuthAdjust}\,(b) shows adjusted $\alpha'_2$ versus the geometric phase $\gamma$: It is clearly seen, that adjusted
$\alpha'_2$ fulfills the theory condition (solid line - color: green line) namely a linear dependency as expressed in Eq.(\ref{eq:Cond3}).

%
%
%
%
%
%
%
%

\section{\label{sec:discussion}Discussion}

If no corrections are applied to the Bell angles the S-value decreases from $2\sqrt{2}$ at $\gamma=0$ to zero at $\gamma=\pi$ and regains the
value of $2\sqrt{2}$ at $\gamma=2\pi$ (Fig.\,\ref{fig:ResultsAzimuthAdjust}\,(a), \emph{S-value without adjustment}). Keeping the azimuthal
angles fixed, an appropriate adjustment of the polar Bell angles determined by the geometric phase $\big(\beta_1=\arctan(\cos\gamma)\big)$,
yields a sinusoidal oscillation of the S-value ($2\le S\le 2\sqrt{2}$, with period $\pi$, see Fig.\,\ref{fig:ResultsPolAdjust}(a)). Finally, the
maximum S-value of $2\sqrt{2}$ can be observed, for all values of the geometric phase  $\gamma$, if the difference of the azimuthal angles
(angle between the analysis planes) equals the geometric phase ($\alpha'_2=\gamma$), while the polar Bell angles remain unchanged at the typical
values for testing of a BI (Fig.\,\ref{fig:ResultsAzimuthAdjust}(a), \emph{S-value azimuthal adjusted}).

In our experiment, unlike the proposed setup in \cite{Durstberger04}, the geometric phase is not acquired solely in one arm of the
interferometer but by two successive rf flippers. Hence, an alternative approach towards  generation of geometric phase is introduced here: The
effect of the rf-flipper inside the interferometer is described by the unitary operator $\hat U(\phi_{\textrm I})$, which induces a spinor
rotation from $\ket{\Uparrow}$ to $\ket{\Downarrow}$, denoted as $\hat U(\phi_{\textrm I})\ket{\Uparrow}=\ket{\Downarrow}$. The rotation axis
encloses an angle $\phi_{\textrm I}$ with the $\hat {\mathbf x}$-direction, and is determined by the oscillating magnetic field
$B^{(1)}=B_{\textrm{rf}}^{(\omega)} \cos(\omega t +\phi_{\textrm{I}})\cdot\hat{\mathbf y}$. Without loss of generality one can insert a unity
operator, given by $\1= \hat U^\dagger(\phi_0)\hat U(\phi_0)$, yielding
\begin{eqnarray}
\hat U(\phi_{\textrm I})\ket{\Uparrow}&=&\overbrace {\hat U(\phi_{\textrm I})\hat U^\dagger(\phi_0)}^{\gamma}\hat
U(\phi_0)\ket{\Uparrow}\hspace{-2.55cm}\underbrace{\hspace{2cm}}_{\1}\\\nonumber&=&e^{i\gamma}\ket{\Downarrow},
\end{eqnarray}
where $\hat U(\phi_0)$ can be interpreted as a rotation from $\ket{\Uparrow}$ to $\ket{\Downarrow}$, with the $\hat {\mathbf x}$-direction being
the rotation axis, and $\hat U^\dagger(\phi_0)$ describes a rotation about the same axis back to the initial state $\ket{\Uparrow}$.
Consequently, $\hat U(\phi_{\textrm I})\hat U^\dagger(\phi_0)$ can be identified to induce the geometric phase $\gamma$, along the reversed
evolution path characterized by $\phi_0$, followed by another path determined by $\phi_{\textrm I}$.

Due to the inherent phase instability of the neutron interferometer, it is necessary to perform a reference measurement for each setting of
$\gamma $ and $\delta$. This is achieved by turning off the rf-flipper inside the interferometer, yielding a reference interferogram. The
oscillations plotted in Fig.\,\ref{fig:OsciPolAdjust} and Fig.\,\ref{fig:OsciAzimuthAdjust} are normalized, by the contrast of the reference
measurement, and the phase of the reference interferogram is taken into account (relative phase between the oscillations).

At this point it should be noted that the average contrast of $\sim$\,50\,\% (obtained for $\delta=\pi/2$ with maximum intensity of $\sim$\,25
neutrons/sec.) is below the threshold of 70.7\,\%, required to observe a violation of a BI.

Violation of a BI, for a spin-path entanglement in neutron interferometry, has already been reported in \cite{Hasegawa03Bell}, the argument here
is the influence of the geometric phase on the $S$-value. Consequently a normalization as performed does not influence the validity of the
results presented here.

Finally we want to discuss some systematic errors in our experiment, in particular in the state preparation and in the projective spin
measurement. Under ideal conditions no interference fringes should be obtained in the H-beam, due to orthogonal spin states in the interfering
sub-beams. Nevertheless we have observed intensity modulations with a contrast of a few per cent. This indicates, that the state preparation (rf
flipper) was not perfect in some sense. The expectation values for the joint measurements
Eqs.(\ref{eq:jointMeasurement})\,-\,(\ref{eq:S-functionAzFixed}) can be deduced for an arbitrary (spin) state, in the path of the IFM where the
rf-flipper is located,
\begin{eqnarray}
\begin{split}
\ket{\Psi_{\textrm{Meas.}}(\gamma)}&=\frac{1}{\sqrt{2}}\Big(\ket{\textrm{I}}\otimes  \ket{\Uparrow}+ e^{i\chi}\ket{\textrm{II}}\otimes
e^{i\gamma}\big(\sin\frac{\theta}{2}\ket{\Uparrow}\\&+\cos\frac{\theta}{2}\ket{\Downarrow}\big)\Big).
\end{split}
\end{eqnarray}
Here $\theta$ is determined by the fringe contrast in the H-beam. These systematic deviations from the theoretical initial state have been taken
into account in the calculation of the final $S$ value.

The asymmetry in the curve of the projective measurement along the $\pm\,\hat{\mathbf x}$ direction of the path measurement, denoted as $\hat
P^{\textrm{p}}_{+x}:({\alpha'_1=\pi/2, \alpha'_2=0})$ and $\hat P^{\textrm{p}}_{-x}:({\alpha'^{\bot}_1=3\pi/2, \alpha'_2=0})$ in
Fig.\,\ref{fig:OsciPolAdjust} is considered to result from a misalignment of the static magnetic fields, at the position of the coil, such as
the stray field of the first guide field, the second guide field and the two fields ($\hat{\mathbf x},~\hat{\mathbf z}$)-direction produces by
the coil itself.

\section{\label{sec:conclusion}Conclusion}

We have demonstrated a technique to balance the influence of the geometric phase, generated by one subspace of the system, considering a BI.
This is achieved by an appropriate adjustment of the polar Bell angles (keeping the measurement planes fixed) or one azimuthal angle (keeping
the polar Bell angles at the well-known values), determined by a laborious measurement procedure. It is demonstrated in particular, that a
geometric phase in one subspace does not lead to a loss of entanglement, determined by a violation of a BI. The experimental data are in good
agrement with theoretical predictions presented in \cite{Durstberger04}, demonstrating the correctness of the procedure as a matter of
principle.

\begin{acknowledgments}
We thank E. Balcar for a critical reading of the manuscript. This work has been supported by the Austrian Science Foundation, FWF (P21193-N20,
T389-N16 and F1513). K.D.-R. would like to thank the FWF for funding her work by a Hertha Firnberg Position.

\end{acknowledgments}

\end{document}